\documentclass[fleqn,usenatbib]{mnras}
\usepackage{amsmath}
\usepackage{graphicx}
\usepackage{physics}
\usepackage{color}
\usepackage{hyperref}
\usepackage[normalem]{ulem}
\usepackage{bm}
\usepackage{footmisc}

\usepackage{booktabs}
\usepackage{overpic}

\newlength{\abovecaptionskip}
\usepackage{threeparttable}

\title[FDM Constraints From HFF]{Fuzzy Dark Matter Constraints from the Hubble Frontier Fields}

\newcommand{\dif}{\text{d}}
\defcitealias{Sipple24}{SL}
\defcitealias{Bouwens21}{B21}
\defcitealias{Bouwens22}{B22}

\usepackage[T1]{fontenc}
\usepackage{ae,aecompl}
\usepackage{newtxtext,newtxmath}

\author[Sipple et al.]{
Jackson Sipple\href{https://orcid.org/0009-0009-4041-8843}{\includegraphics[width=1em,height=1em]{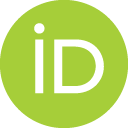}},$^{1}$\thanks{Contact e-mail: \href{mailto:jsipple@sas.upenn.edu}{jsipple@sas.upenn.edu}}
Adam Lidz\href{https://orcid.org/0000-0002-3950-9598}{\includegraphics[width=1em,height=1em]{orcid-ID.png}},$^{1}$
Daniel Grin\href{https://orcid.org/0000-0002-5084-8961}{\includegraphics[width=1em,height=1em]{orcid-ID.png}}$^{2}$
and Guochao Sun\href{https://orcid.org/0000-0003-4070-497X}{\includegraphics[width=1em,height=1em]{orcid-ID.png}}$^{3}$
\\
$^{1}$Center for Particle Cosmology, Department of Physics and Astronomy, University of Pennsylvania, Philadelphia, PA 19104, USA\\
$^{2}$Department of Physics and Astronomy, Haverford College, PA 19041, USA\\
$^{3}$CIERA and Department of Physics and Astronomy, Northwestern University, IL 60201, USA
}

\date{Accepted XXX. Received YYY; in original form ZZZ}
 
\pubyear{\the\year{}}

\begin{document}
\label{firstpage}
\pagerange{\pageref{firstpage}--\pageref{lastpage}}
\maketitle

\begin{abstract}

In fuzzy dark matter (FDM) cosmologies, the dark matter consists of ultralight bosons ($m\lesssim10^{-20}$ eV). The astrophysically large de Broglie wavelengths of such particles hinder the formation of low-mass dark matter halos. This implies a testable prediction: a corresponding suppression in the faint-end of the ultraviolet luminosity function (UVLF) of galaxies. Notably, recent estimates of the faint-end UVLF at $z\sim5-9$ in the Hubble Frontier Fields, behind foreground lensing clusters, probe up to five magnitudes fainter than typical (``blank-field") regions. These measurements thus far disfavor prominent turnovers in the UVLF at low luminosity, implying bounds on FDM. We fit a semi-empirical model to these and blank-field UVLF data, including the FDM particle mass as a free parameter. This fit excludes cases where the dark matter is entirely a boson of mass $m<1.5\times10^{-21}$ eV (with $2\sigma$ confidence). We also present a less stringent bound deriving solely from the requirement that the total observed abundance of galaxies, integrated over all luminosities, must not exceed the total halo abundance in FDM. This more model-agnostic bound disfavors $m<5\times10^{-22}$ eV ($2\sigma$). We forecast that future UVLF measurements from JWST lensing fields may probe masses several times larger than these bounds, although we demonstrate this is subject to theoretical uncertainties in modeling the FDM halo mass function.

\end{abstract}

\setlength{\footnotemargin}{\parindent}

\section{Introduction}
\label{sec:intro}
While there is much evidence the majority of matter in the universe is {\em dark} (i.e. does not interact appreciably with electromagnetic fields) \citep{Bertone18b,Planck20,Arbey21}, its precise character remains elusive \citep{Bertone18,cirelli24}. In this work we assess the possibility that the dark matter consists of fuzzy dark matter (FDM) particles \citep{Hu_2000}, ultralight bosons with particle masses of $m\sim 10^{-22}-10^{-20}$ eV).\footnote{This is the range of interest for our present analysis although the term  may describe several decades in mass around this range.}
FDM is sometimes alternatively referred to as wave, axion-like, or Bose-Einstein condensate dark matter, as it is a subclass of these categories. 
The cosmological de Broglie wavelengths in FDM leave detectable signatures in the large-scale structure of the universe (see e.g. \citealt{Marsh16b,Schwabe18,Niemeyer20,Hui21,Ferreira21,Ohare24} for reviews). We contrast this with conventional cold dark matter (CDM): a collisionless, non-relativistic species without wave-like effects on astrophysical scales. 

Early work (e.g. \citealt{Hu_2000}, see \citealt{Lee18} for a brief history) often considered FDM as a resolution of the so-called ``small-scale'' problems of CDM (e.g., the missing satellite, cuspy halo, and too-big-to-fail problems, see \citet{Popolo17,Bullock17} for reviews). While it is possible that that these issues owe partly to observational incompleteness \citep{Kim18} and/or may be resolved through including the impact of baryonic feedback in CDM \citep{Sawala16}, it is important to recognize that CDM is less-sharply tested on small spatial scales. Alternative possibilities such as FDM make distinctive small-scale predictions and thereby provide a useful foil for CDM models. It may be that FDM is what is realized in nature.
Indeed, FDM is well-motivated from particle physics considerations: ultralight scalar particles with the correct relic density to make up the dark matter \citep{Marsh16b,Hui17,Hui21}  arise naturally in string theory \citep{Arvanitaki10} or owing to symmetry-breaking processes in the early universe, e.g. via the misalignment mechanism analogous to that of the proposed QCD axion \citep{Peccei77,Wilczek78,Weinberg78}. 
There are additional well-motivated alternatives to CDM besides FDM, including warm dark matter (WDM) in which free-streaming leads to a suppression of small scale fluctuations. Although our focus in this work is on FDM, we will briefly consider the implications of our results for WDM as well. 

The ultraviolet luminosity function (UVLF) describes the number density of galaxies as a function of their luminosity. The form of the UVLF depends on the underlying abundance of dark matter halos and how galaxy luminosity scales with host halo mass \citep{Yang03,Cooray05,Bouwens15,Schive_2016,Sipple24}. Therefore it is possible to use the UVLF to test theories, like FDM, which predict small dark matter halos to be rare. Analyses using the gravitational lensing of galaxies in the Hubble Frontier Fields (HFF) \citep{Coe15,Lotz17} have probed the UVLF at $z\leq9$ at up to five magnitudes fainter \citep{Livermore17,Ishigaki18,Atek18,Bhatawdekar19,Bouwens22} than in unmagnified regions (known as the \textit{field} or \textit{blank-field} regions) \citep{Bouwens21}. 

Lensed UVLF measurements are inherently challenging since magnification factor and survey volume estimates
require modeling cluster mass distributions. Indeed, if unaccounted systematic errors remain in the UVLF measurements at low luminosity, this could influence our constraints. 
In this work we assume the binned UVLF measurements of \citet{Bouwens22} and refer the reader to that work and its companion, \citet{Bouwens22_I}, for a more detailed discussion of potential uncertainties in determining the UVLF behind lensing clusters and progress in mitigating these effects. For instance, they marginalize over uncertainties in the cluster mass distribution model and reassuringly find consistency between the faint-end slope estimates from the field and behind the cluster lenses. Conservatively, we also include separate bounds using only the \citet{Bouwens21} field measurements.

The faintest galaxies in the lensed samples reside in halos that are roughly an order of magnitude smaller in mass than any in the field \citep{Sipple24}, yet the UVLF data do not show an obvious faint-end suppression.
As we will see, these measurements thereby tighten constraints on the FDM particle mass compared to previous UVLF studies \citep{Schive_2016,Corasaniti17,Menci_2017,Ni_2019,Lapi22,Winch24} making it competitive with similar bounds from observations of the Lyman-$\alpha$ forest \citep{Irsic17,Kobayashi17,Armengaud17,Rogers21} and Milky Way satellite galaxies \citep{Schutz20,Nadler21,Banik21}. The consistency across multiple independent probes helps establish the robustness of the FDM constraints. 

Furthermore, JWST has already begun conducting revolutionary measurements of the bright-end UVLF at $z\gtrsim10$ \citep{Harikane23,Bouwens23,Harikane24,Donnan24}. This has produced a surprising number of bright photometric galaxy candidates, with some spectroscopic confirmations, even from when the universe was only a few hundred Myr old. However, the current parameter space in FDM particle mass allows deviations from CDM predictions only in the smallest, faintest galaxies. JWST has not yet delivered results in this regime, as this requires ongoing and future surveys behind foreground galaxy cluster lenses. Our work aims to provide the current state-of-the-art UVLF constraints on FDM in anticipation that these same methods can be applied to upcoming JWST measurements.

This paper is organized as follows: In Section \ref{sec:method} we describe our approach for modeling the UVLF based on halo mass function models in the literature and empirically-calibrated relationships between host dark matter halo mass and galaxy UV luminosity. In Section \ref{sec:results} we present our results including constraints on the FDM particle mass. In Sections \ref{sec:LM} and \ref{sec:hmf} we consider the robustness of these constraints to uncertainties in how UV luminous galaxies populate their host halos, and in modeling the halo mass function in FDM, respectively. We then put our new FDM mass bounds in context in Section \ref{sec:constraints} by comparing them with constraints from previous UVLF studies and with results from other techniques. Looking forward, we discuss the prospects for future improvements using upcoming JWST data in Section \ref{sec:jwst}.  In Section \ref{sec:conclusion} we present our conclusions and also briefly discuss the implications of our results for other alternatives to CDM.
Throughout, we assume the following cosmological parameters: $\Omega_m=0.31$, $\Omega_\Lambda=0.69$, $\Omega_b=0.049$, $h=0.677$, $\sigma_8=0.818$, $n_s=0.96$ in agreement with Planck 2018 determinations \citep{Planck20}.

\section{Methodology}
\label{sec:method}
Our methodology closely follows the approach taken in \citet{Sipple24}, except we now consider FDM models, along with CDM, and include the FDM particle mass as a new parameter.  
In this section we first describe how we parametrize the effect of the FDM particle mass on the halo mass function (HMF) (Section \ref{sec:method-hmf}). Then we summarize our modified UVLF modelling procedure (Section \ref{sec:method-uvlf}).

\subsection{The Halo Mass Function in FDM}
\label{sec:method-hmf}
\begin{figure}
  \centering
  \includegraphics[width=0.45\textwidth]{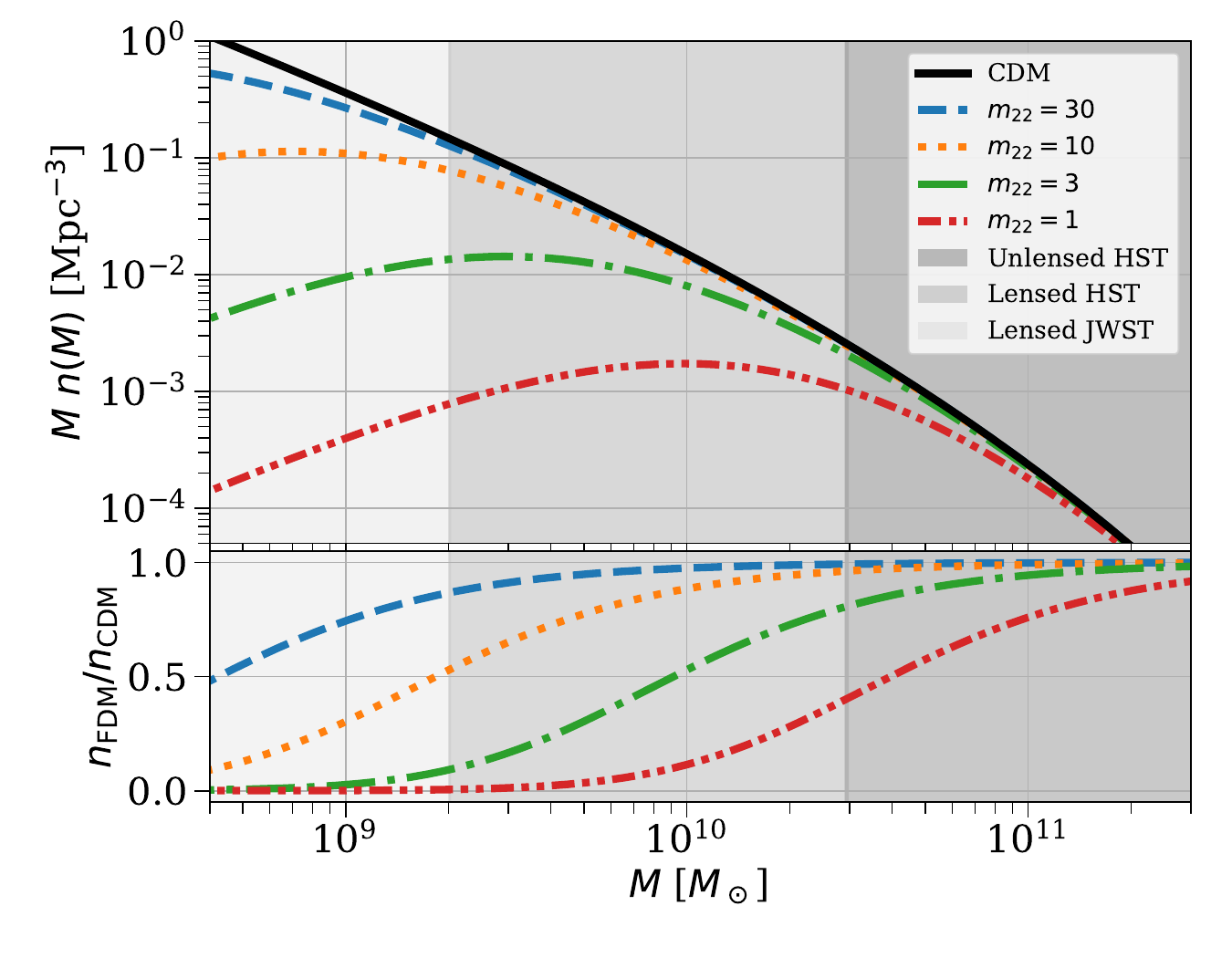}
  \caption{{\em Top:} The FDM halo mass function fit described by Equation \ref{eq:schive} \citep{Schive_2016} at $z=8$. The halo mass function, as plotted here, describes the co-moving abundance of halos per logarithmic interval in mass. 
  The black solid curve shows the CDM prediction, while the broken colored lines represent various FDM particle masses ($m_{22}=m_{\rm FDM}/10^{-22}$eV). The shaded regions delineate the approximate limiting halo mass scales probed by HST data sets (\citetalias{Bouwens21}, \citetalias{Bouwens22}) \citep{Sipple24} and JWST projections (Section \ref{sec:jwst}). {\em Bottom:} The ratio of each FDM mass function to the CDM HMF.}
  \label{fig:dndM}
\end{figure}
If the dark matter is primarily FDM its large scale evolution matches CDM, but structure formation near and below its de Broglie wavelength is opposed by the uncertainty principle. The balance between this effect and gravity sets a relevant Jeans scale, $k_J$, where there is significant deviation from CDM for $k\gtrsim k_J$ \citep{Khlopov85,Hu_2000}. In linear theory this co-moving Jeans wavenumber scales as $k_J\propto(1+z)^{-1/4}$ during matter domination \citep{Hu_2000}, i.e. FDM suppression effects apply to smaller co-moving scales over time. 
Consequently, it is common to approximate FDM as evolving like CDM with a suppression in the {\em initial linear power spectrum} (at matter-radiation equality), ignoring subsequent dynamical effects which would modify the growth of the smaller scale modes. In most of this work we employ fitting formulas for the halo mass function derived from FDM simulations that use this approximation, as full FDM simulations in large cosmological volumes require a currently prohibitive dynamic range in spatial scale.
In Section \ref{sec:hmf} we return to this point and consider the sensitivity of our results to this approximation.

More concretely, the initial power spectrum is suppressed by a factor of more than one-half above $k_{1/2}\sim 4.5m_{22}^{4/9}$ Mpc$^{-1}\sim 0.5k_J$  \citep{Hu_2000}, where $m_{22}=m_{\rm FDM}/10^{-22}$eV.\footnote{Some sources (e.g. \citealt{Schive_2016}) define $k_{1/2}$ instead as where the \textit{transfer function} equals $1/2$ (and the power is $1/4$ the CDM value).} 

The halo mass function describes the number density of halos per mass interval. As structure grows from the initial density perturbations, the formation of halos near and below the half-mode mass scale $M_{1/2}=(4\pi\bar\rho_m/3)(\pi/k_{1/2})^3 = 5.6\times10^{10}m_{22}^{-4/3} M_\odot$ will be suppressed. The FDM dynamical effects mentioned above, including complex non-linear wave effects, may also influence the shape and redshift evolution of the HMF (see e.g. \citealt{Hui17,Ferreira21}). Current studies suggest that dynamical effects become important on scales where structure is already heavily suppressed \citep{Zhang18,Nori19,Li19,May23}, as alluded to above, and so they are often neglected in FDM HMF models.

In this work we primarily consider the HMF of \citet{Schive_2016}, which proposes a fitting formula to simulations with FDM particle masses around $m_{22}\sim1$. These simulations assume CDM-like growth with an FDM initial matter power spectrum. More recent simulations have found similar results (\citealt{Corasaniti17,Nori19,Ni_2019,May23}, although see Section \ref{sec:hmf} for further discussion regarding alternate halo mass function models).
Adopting the Sheth-Tormen halo mass function in CDM, $n_{\rm CDM}(M,z)$ \citep{Sheth:2001dp}, \citet{Schive_2016} find that their simulated FDM mass-function, $n_{\rm FDM}(M,z)$ is well-described by:
\begin{equation}
    n_{\rm FDM}(M,z) \simeq n_{\rm CDM}(M,z) \left[1+\left(\frac{M}{M_0}\right)^{-1.1}\right]^{-2.2},
    \label{eq:schive}
\end{equation}
with $M_0=1.6\times10^{10}m_{22}^{-4/3}M_\odot\approx0.3M_{1/2}$. This matches CDM for $M>>M_0$,  but drops to $\sim80\%/20\%/0.5\%$ the CDM value for $M=10M_0,M_0,0.1M_0$.  
The halo suppression mass scale shrinks with increasing particle mass as $M_0\propto m_{22}^{-4/3}$. Qualitatively, these effects can be seen as a consequence of the uncertainty principle. 
As a larger particle mass implies a smaller de Broglie wavelength $\lambda_{\rm dB}\propto 1/m$) it is possible to confine the dark matter within smaller regions. In the context of the HMF, heavier FDM particles are more difficult to distinguish from CDM. In this parametrization CDM corresponds to the $m_{22}\rightarrow\infty$ limit. 

We compare the CDM and FDM mass functions at $z=8$ in Figure \ref{fig:dndM} for the cases of $m_{22}=1,3,10,$ and $30$. The top panel shows the mass functions themselves while the bottom gives the ratio between each FDM model and CDM. The gray shaded regions illustrate the approximate mass scales probed by different UVLF measurement sets assuming the best-fit luminosity-halo mass relation of \citet{Sipple24} (or, as we find equivalently, this work (Section \ref{sec:results})). The shaded regions, from dark to light, show: the field (unlensed) HST \citepalias{Bouwens21} ($\gtrsim 3\times10^{10}M_\odot$), lensed HST \citepalias{Bouwens22} ($\gtrsim2\times10^9M_\odot$), and forecasted lensed JWST (Section \ref{sec:jwst}) \citep{Jaacks19,Shen23} ($\gtrsim4\times10^8M_\odot$). Observational probes which investigate the HMF at lower mass scales will be better able to distinguish between FDM scenarios and CDM. Equation \ref{eq:schive} gives a redshift-independent ratio between the HMF in FDM and CDM. However, plausible alternative FDM halo mass function models predict more pronounced differences with CDM at higher redshift  (see Section \ref{sec:hmf}).

\subsection{UVLF Modelling}
\label{sec:method-uvlf}
Our UVLF modelling procedure is similar to \citet{Sipple24}. The dataset we consider is a joint set of UVLF measurements at redshifts $z=5-10$ observed by HST and analyzed by \citet{Bouwens21} \citepalias{Bouwens21} (with $z=10$ determinations from \citealt{Oesch18}) and \citet{Bouwens22} \citepalias{Bouwens22}. The latter exploits gravitational lensing magnification from foreground galaxy clusters in the Hubble Frontier Fields (HFF) to probe the UVLF at luminosities up to five magnitudes fainter than the field alone. We apply a correction for dust attenuation described in \citet{Sipple24} based on previous treatments \citep{Vogelsberger20,Smit12}, although this effect is negligible at the faint end ($M_{\rm UV}\gtrsim-17$).

\subsubsection{The Conditional Luminosity Function}
We follow previous work \citep{Yang03,Cooray05,Bouwens15,Schive_2016,Sipple24} and parameterize the median galaxy UV luminosity, $L_c(M,z)$, as a double power law in host halo mass, $M$, and a single power law in redshift, $z$,
\begin{equation}
    L_c(M,z) = L_0\left[\frac{(M/M_1)^p}{1+(M/M_1)^q}\right]\left(\frac{1+z}{7}\right)^r, \label{eq:L_c}
\end{equation}
where the five parameters, $L_0$, $M_1$, $p$, $q$, and $r$, are calibrated using UVLF measurements. The double power-law form reflects the expectation that feedback effects suppress star formation (and hence UV luminosities) most strongly in low-mass and high-mass halos (see \citet{Silk11} for a brief review).\footnote{Other approaches, such as \texttt{GALLUMI} \citep{Sabti22}, instead model the ratio of stellar mass to halo mass (or the ratio of their time derivatives) as a double power-law in halo mass. Such models should be comparable to the conditional luminosity function approach used in this paper, as UV luminosity and star formation rate are well-correlated and proportional to one another \citep{Madau14} (see e.g. \citealt{Sipple24} for a discussion on the connection of the conditional luminosity function to other physical quantities).} We explore the impact of uncertainties in the role of feedback effects in Section \ref{sec:LM}.

The conditional distribution connecting the UVLF and HMF, $\phi_c(L|M,z)$, is taken to be a lognormal with scatter $\sigma_{\rm LN}=0.37$ and median $\mu_{\rm LN}=L_c(M,z)$.\footnote{We fix the value of the scatter following previous treatments \citep{Sipple24,Schive_2016,Bouwens15,Bouwens08} as it is generally not well-constrained.} Assuming each dark matter halo above some limiting mass $M_{\rm min}$ will, on average, host one luminous galaxy (supported by low satellite fractions at $z\sim6$ \citep{Harikane22}) and taking a duty cycle of unity, the relation to the UVLF $\phi(L,z)$ is:
\begin{equation}
    \phi(L,z) = \int_{M_{\rm min}}^\infty  \phi_c(L|M,z) n(M, z) \dif M. \label{eq:phi}
\end{equation}
Unlike \citet{Sipple24}, here $n(M, z)$ represents the FDM halo mass function of Equation \ref{eq:schive}, which is parameterized by a particle mass $m_{22}$. As mentioned earlier, CDM scenarios are captured by the large $m_{22}$ limit of this mass function.

Explicitly, the vector of parameters of our model fit is defined as ${\bm \theta} = \left(p,q,r,M_{\rm UV,0},{\rm log_{10}}(M_1/{\rm M}_\odot), 1/m_{22}\right)$, where $M_{\rm UV,0}$ is $L_0$ in units of absolute magnitude \citep{Oke74}\footnote{$M_{\rm UV} = 51.6 - 2.5\log_{10}\left(L/[\text{erg} \, \text{s}^{-1}\text{Hz}^{-1}]\right)$. Note that under a change of variables: $\phi(L)\text{d}L=\phi(M_{\rm UV})\text{d}M_{\rm UV}$.}. Our priors on these parameters are: ([0.8, 3.5], [0.8, 2.5], [0, 2.5], [-25, -20], [10, 14], [0, 2]). The prior ranges for the first five parameters are chosen based on previous fits using this parametrization of $L_c(M)$ under CDM, especially those in \citet{Sipple24}. 
Ultimately, the likelihood functions are well-peaked within the range spanned by these priors. The upper bound on the $1/m_{22}$ prior is chosen to be conservative as $m_{22}\lesssim1$ has been excluded by previous UVLF modeling studies at $>2\sigma$ \citep{Schive_2016,Corasaniti17}. We choose to sample in inverse $m_{22}$ to allow the CDM limit $m_{22}\rightarrow\infty$ to be explicitly considered in the prior volume (where $1/m_{22}\rightarrow0)$. See Section \ref{sec:LM-prior} for additional discussion regarding the FDM mass prior. Likelihoods are determined following \citet{Sipple24}, where the likelihood of a model UVLF value at a given magnitude is Gaussian where symmetric errors are quoted, half-Gaussian when only upper limits are quoted, or follows the prescription of Equations 13-16 of \citet{Barlow04} for asymmetric errors. 

We use the Markov chain Monte Carlo (MCMC) sampler from the \texttt{emcee} python package \citet{Foreman-Mackey12} to sample the posterior distribution $p(\mathbf{\theta})$ and obtain confidence intervals and correlations between parameters.\footnote{In each MCMC run in this work we initialize 96 walkers in the prior volume and run many times more iterations ($\sim10,000$) than the longest autocorrelation time of any parameter ($\sim100$) to ensure convergence and stability. We further remove the first few thousand samples to remove influence from our initialization.}

\section{Results}
\label{sec:results}
\begin{figure*}
  \centering
  \includegraphics[width=\textwidth]{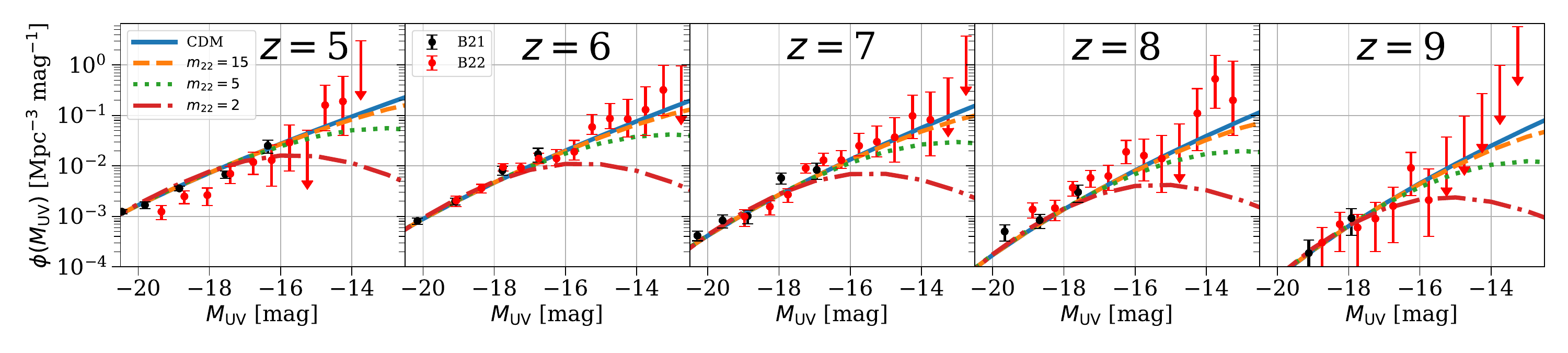}
  \caption{Faint-end $z=5-9$ UVLF measurements from \citetalias{Bouwens21} (field galaxies) and \citetalias{Bouwens22} (lensed galaxies) with 1$\sigma$ error bars (or upper limits) compared with best-fit model UVLFs for CDM and FDM with fixed $m_{22}=15$ (orange dashed), $m_{22}=5$ (green dotted), or $m_{22}=2$ (red dot-dashed). The model fits use the full \citetalias{Bouwens21}+\citetalias{Bouwens22} set of measurements although we show here only the UVLF at relatively low luminosities, where FDM has the most impact, for illustration.}
  \label{fig:UVLF}
\end{figure*}

Figure \ref{fig:UVLF} shows best fit UVLF curves under CDM and different fixed values of $m_{22}$ at $z=5-9$ compared to \citetalias{Bouwens21} and \citetalias{Bouwens22} measurements. 
The first thing to note is the CDM model provides a better fit in each redshift bin than the FDM models shown. The UVLF data do not show any turnover at low luminosities which might provide evidence of FDM over CDM.
The best-fit CDM model (Table \ref{table:params}) is a reasonably good description of the data, with a reduced chi-squared value of $\chi^2_\nu = 1.69$ for a fit with 106 degrees-of-freedom (111 data points and five model parameters). 

Given the lack of a low-luminosity turnover, the data can be used to bound the FDM particle mass. 
Specifically, the FDM mass controls the scale at which suppression in the mass function becomes significant, where $M\lesssim M_0\propto m^{-4/3}_{22}$. This leads to a strong suppression of the faint-end UVLF due to a relative dearth of galaxies of luminosity $L<L_c(M_0, z)$, without affecting the rest of the UVLF. Furthermore, there is little way to fit both the suppressed and non-suppressed regimes by tuning the other model parameters. 
Therefore, FDM scenarios which predict a suppression in this regime, with $m_{22}\sim\mathcal{O}(1)$ and below, are strongly excluded by the data. Quantitatively, allowing the FDM particle mass to vary, we find $m_{22}<15.1$ disfavored at $2\sigma$ (Tables \ref{table:params},\ref{table:m22}).

\begin{table*}
\begin{threeparttable}
\caption{\label{table:params}Best-fit UVLF model parameters and 68\% confidence intervals determined from MCMC sampling under different modelling assumptions. The first row displays the main results of our fiducial model, described in Sections \ref{sec:method} and \ref{sec:results}. The next two rows show results when restricted to CDM-only scenarios or to not use the lensed UVLF measurements from \citetalias{Bouwens22} respectively. The following three rows are modified models described in Section \ref{sec:LM}: adding additional parameters to Equation \ref{eq:L_c} (Section \ref{sec:LM-tpl}) or with alternate priors for $m_{22}$ (Section \ref{sec:LM-prior}) of either flat in $\log_{10}(m_{22})$ or based on the Jeffreys prior. The final two rows use the alternative FDM HMF parametrizations of \citet{Kulkarni21} (Section \ref{sec:hmf-sk}) or \citet{Marsh13,Marsh16} (Section \ref{sec:hmf-marsh}).}
\begin{tabular}{@{}l*{6}{c}@{}}
\toprule
Model      & $p$  & $q$  & $r$  & $M_{\text{UV},0}$  & $\log_{10} \left(M_1/M_\odot\right)$ & $1/m_{22}$ \\ \midrule
Fiducial (\S\ref{sec:method},\ref{sec:results}) & $1.69\pm^{0.03}_{0.03}$ & $1.57\pm^{0.18}_{0.15}$ & $1.12\pm^{0.11}_{0.11}$ & $-23.79\pm^{0.19}_{0.18}$ & $11.93\pm^{0.06}_{0.06}$ & $1.76\pm^{2.07}_{1.23}\times10^{-2}$ 
\\
\midrule
CDM Only     & $1.70\pm^{0.03}_{0.03}$ & $1.57\pm^{0.18}_{0.15}$ & $1.12\pm^{0.11}_{0.11}$ & $-23.78\pm^{0.19}_{0.18}$ & $11.93\pm^{0.06}_{0.06}$ & $-$ 
\\
Blank-Fields \citepalias{Bouwens21} Only & $1.80\pm^{0.09}_{0.08}$ & $1.33\pm^{0.13}_{0.11}$ & $0.85\pm^{0.13}_{0.13}$ & $-23.11\pm^{0.46}_{0.38}$ & $11.69\pm^{0.13}_{0.15}$ & $1.19\pm^{1.42}_{0.85}\times10^{-1}$ 
\\
\midrule
Triple Power Law (\S\ref{sec:LM-tpl})\tnote{$\dagger$}  & $2.08\pm^{0.13}_{0.11}$ & $1.42\pm^{0.08}_{0.08}$ & $1.12\pm^{0.11}_{0.11}$ & $-22.37\pm^{0.48}_{0.43}$ & $11.44\pm^{0.13}_{0.13}$ & $6.64\pm^{5.17}_{4.27}\times10^{-2}$ 
\\
Log Prior (\S\ref{sec:LM-prior}) & $1.69\pm^{0.03}_{0.03}$ & $1.58\pm^{0.18}_{0.15}$ & $1.12\pm^{0.11}_{0.11}$ & $-23.79\pm^{0.18}_{0.18}$ & $11.93\pm^{0.06}_{0.06}$ & $1.84\pm^{1.42}_{0.63}\times10^{-2}$
\\
Jeffreys Prior (\S\ref{sec:LM-prior}) & $1.69\pm^{0.03}_{0.03}$ & $1.58\pm^{0.18}_{0.15}$ & $1.12\pm^{0.11}_{0.11}$ & $-23.80\pm^{0.18}_{0.18}$ & $11.93\pm^{0.06}_{0.06}$ & $2.53\pm^{2.17}_{1.49}\times10^{-2}$
\\
\midrule
Sharp-$k$ HMF (\S\ref{sec:hmf-sk}) & $1.69\pm^{0.03}_{0.03}$ & $1.55\pm^{0.17}_{0.14}$ & $1.49\pm^{0.11}_{0.11}$ & $-23.77\pm^{0.18}_{0.18}$ & $11.82\pm^{0.06}_{0.06}$ & $4.47\pm^{3.84}_{2.98}\times10^{-2}$
\\
Marsh HMF (\S\ref{sec:hmf-marsh}) & $1.69\pm^{0.03}_{0.03}$ & $1.57\pm^{0.18}_{0.15}$ & $1.15\pm^{0.11}_{0.11}$ & $-23.80\pm^{0.18}_{0.18}$ & $11.93\pm^{0.06}_{0.06}$ & $1.57\pm^{2.24}_{1.16}\times10^{-2}$
 \\ \bottomrule
\end{tabular}
\begin{tablenotes}
       \item[$\dagger$] Additional parameters: $\log_{10}\left(M_2/M_\odot\right)=10.55\pm^{0.05}_{0.05}$, $p_2=1.36\pm^{0.07}_{0.08}$.
\end{tablenotes}

\end{threeparttable}
\end{table*}

\begin{table}
\begin{centering}

\begin{tabular}{@{}lccc@{}}
\toprule
Method & $2\sigma$  & $3\sigma$ & $5\sigma$ 
\\ \midrule
Fiducial (\S\ref{sec:method},\ref{sec:results}) & 15 & 10 & 6.7
\\
No Lensing (\citetalias{Bouwens21} Only) & 2.3 & 1.6 & 1.1
\\
Cumulative UVLF (\S\ref{sec:LM-cumu}) & 5.1 & 4.5 & 3.9
\\
Triple Power Law (\S\ref{sec:LM-tpl}) & 5.8 & 4.4 & 3.2
\\
Log Prior (\S\ref{sec:LM-prior}) & 18 & 11 & 6.5
\\
Jeffreys Prior (\S\ref{sec:LM-prior}) & 13 & 9.4 & 6.0
\\
Sharp-$k$ HMF (\S\ref{sec:hmf-marsh}) & 8.3 & 6.4 & 5.0
\\
Marsh HMF (\S\ref{sec:hmf-marsh}) & 14 & 9.3 & 5.9

 \\ \bottomrule
\end{tabular}  
\end{centering}
\caption{\label{table:m22}Summary of lower bounds on FDM mass (in $m_{22}$) under different model assumptions in this work.
}
\end{table}

\begin{figure}
\setlength{\lineskip}{-3pt}
  \centering
\includegraphics[width=0.45\textwidth]{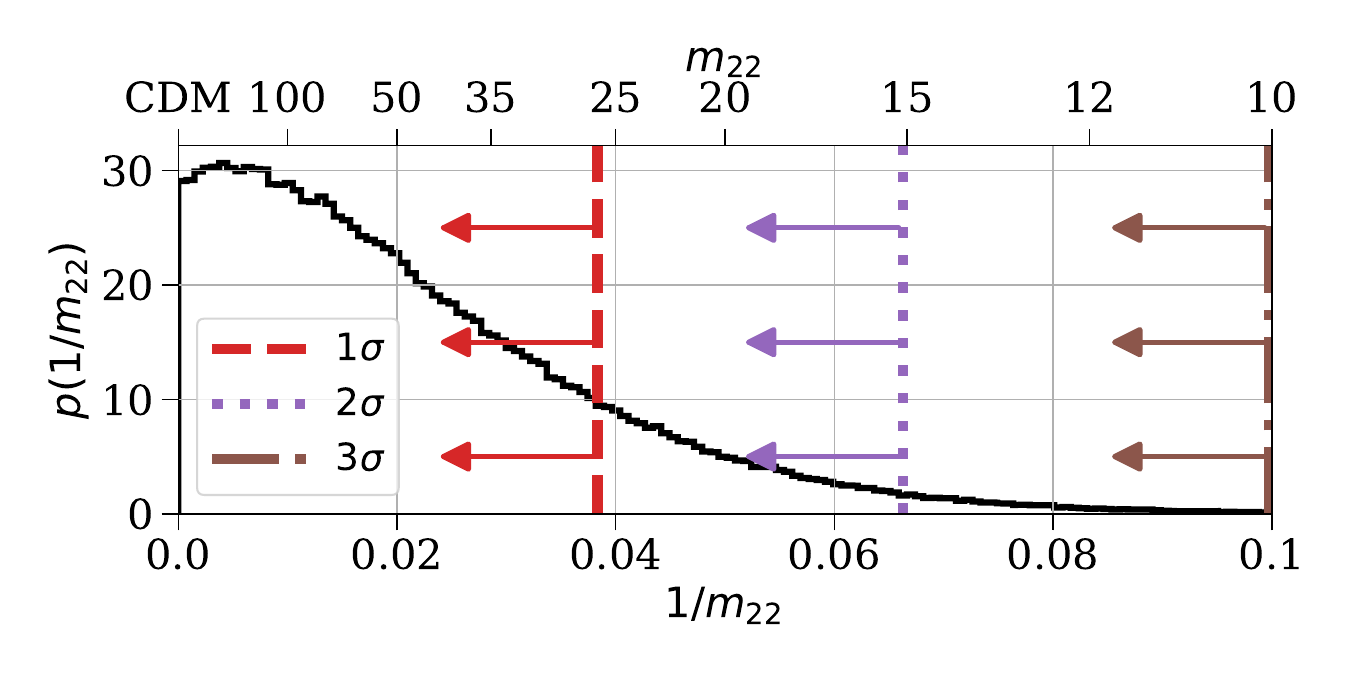}
\includegraphics[width=0.45\textwidth,trim=2 2 2 2,clip]{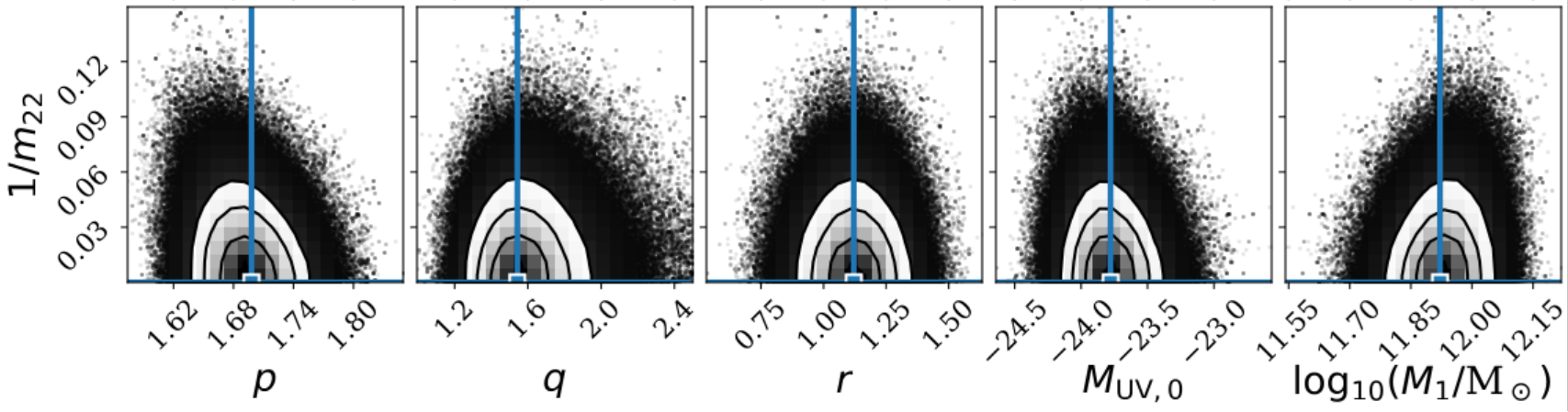}
\caption{\textit{Top:} 1D marginal histogram of FDM mass in our fiducial model from MCMC sampling. The $1/2/3\sigma$ limits are shown as broken vertical lines. Models with $m_{22}\gtrsim100$, including CDM, are indistinguishable from one another using the current data and this region is prior-dominated. \textit{Bottom:} 2D marginal distributions of $1/m_{22}$ with the other five parameters. Blue lines and rectangles show best-fit parameter values from \texttt{scipy.optimize}. There is little covariance with the other parameters, whose distributions differ little from the fit assuming CDM (see Table \ref{table:params} and \citetalias{Sipple24}).
    }
  \label{fig:mcmc}
\end{figure}

Figure \ref{fig:mcmc} shows the 1D posterior for FDM mass, marginalized over the other parameters, and the 2D posterior distributions between $1/m_{22}$ and each of the five parameters defined by Equation \ref{eq:L_c}. The 1D posterior shows that 84\%/98\%/99.9\% of the probability lies above $m_{22}\sim25/15/10$. The data is unable to distinguish between models where $m_{22}\gtrsim100$, which includes the CDM model at $m_{22}\rightarrow\infty$. .

There is almost no covariance between $1/m_{22}$ and the other parameters, with the largest correlation coefficient in magnitude being $\sim-0.16$ with the faint-end slope parameter $p$.\footnote{Correlations between the five parameters of Equation \ref{eq:L_c} are similar to \citet{Sipple24} (see their Figure 2 and Section 4.3).} That is, at least in the context of a double power-law model, the FDM bound is not strongly subject to uncertainties in the relationship between UV luminosity and host halo mass. To illustrate, consider the extreme case of the $m_{22}=2$ curve in Figure \ref{fig:UVLF}. In this model a single power law index covers a wide range in luminosity and so fitting galaxies in the range $-16 \lesssim M_{\rm UV} \lesssim -13$ would necessarily come at the expense of fitting $-20 \lesssim M_{\rm UV} \lesssim -16$ less well. Additionally, while the relative suppression of the HMF in FDM asymptotes to a power law at low mass (see Equation \ref{eq:schive}), its intermediate effect will not be perfectly compensated for by a corresponding power law enhancement in the luminosity-halo mass relation. This further motivates an exploration into our model assumptions, which we undertake in the rest of the paper.

In the following two sections we consider the robustness of these conclusions to a range of alternative assumptions, such as allowing additional freedom in the star formation efficiency of small halos or alternate models of the FDM HMF.

\section{Sensitivity to Model Assumptions}
\label{sec:LM}
The faint end of the UVLF, even including the faint, lensed galaxies of \citetalias{Bouwens22}, is well-fit by the double power-law luminosity-halo mass relation \citep{Sipple24}. This assumption is also consistent with results from zoom-in hydrodynamic simulations of galaxy formation from the FIRE collaboration \citep{Ma18}. Yet, it also may be the case that feedback effects (which arrest star formation) are weaker (e.g. if star formation is exceptionally ``bursty" \citealt{Furlanetto22,Sun23}) at the faint-end, or non-hierarchical structure formation in FDM may lead to younger, brighter stars in the lowest-mass halos \citep{Corasaniti17,Ni_2019,Shen23}.\footnote{The possibility of stronger feedback effects in the smallest halos (e.g. from radiative feedback during the reionization process, \citealt{Borrow23}) is also not captured by this simple model. However, we do not find a clear turnover in the current UVLF data and, unlike the previous scenarios, neglecting this effect in our analyses leads to only a more conservative bound on FDM mass.}

Such possibilities motivate considering alternate approaches for determining the FDM bounds (Section \ref{sec:LM-cumu}) and more flexible models
for the UV luminosity-halo mass relation in FDM (Section \ref{sec:LM-tpl}). First, we place a more conservative constraint using the cumulative UVLF (integrated over all observed luminosities), which assumes little about the UV luminosity-halo mass relationship. Second, we consider extensions beyond our fiducial double power-law model (Equation \ref{eq:L_c})
which allow more freedom in $L_c(M,z)$. Finally, we test the sensitivity of our constraints to our choice of prior on the FDM particle mass (Section \ref{sec:LM-prior}).

\subsection{Cumulative UVLF}
\label{sec:LM-cumu}

\begin{figure*}
  \centering
  \includegraphics[width=\textwidth]{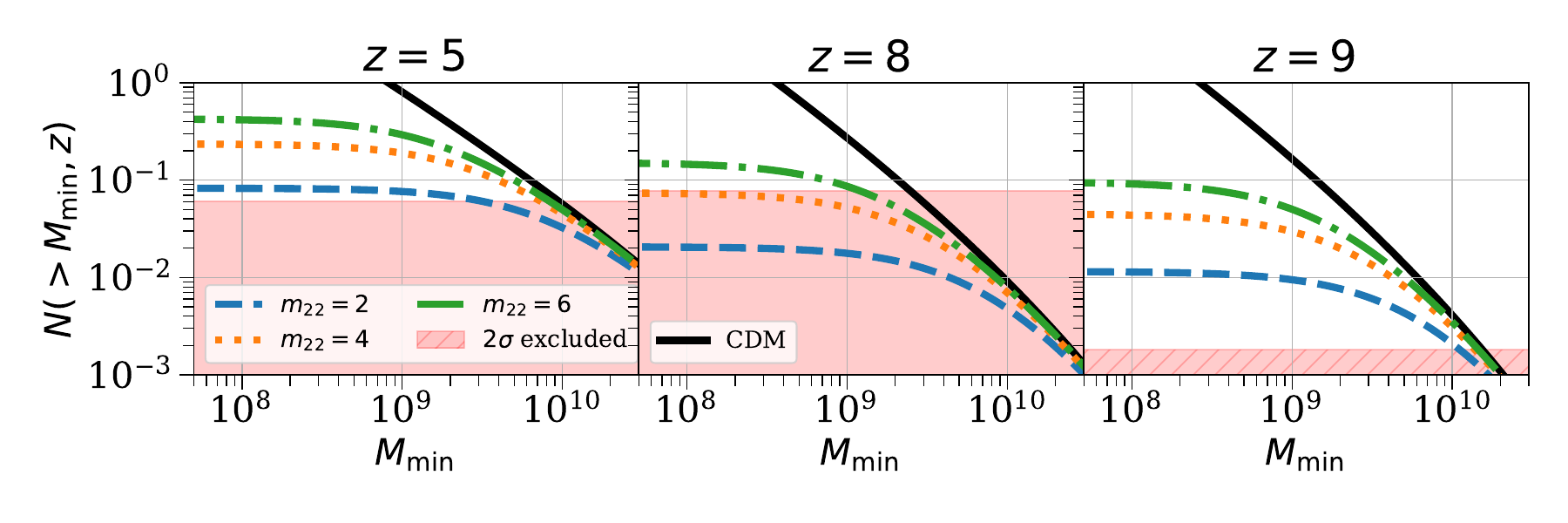}
  \caption{Comparison between the observed cumulative abundance of galaxies and the predicted abundance of dark matter halos as a function of the minimum galaxy-hosting halo mass. Unlike in CDM (black solid), the strong suppression at small halo masses in FDM means the FDM cumulative halo abundance (colorful broken) saturates at some finite value, nearly insensitive to $M_{\rm min} \lesssim 10^8 M_\odot$.
  In viable FDM models, the cumulative halo abundance must exceed the cumulative galaxy abundance. The pink shaded region is $> 2\sigma$ excluded by the cumulative UVLF \citepalias{Bouwens22}.}
  \label{fig:cumulative}
\end{figure*}

The strong suppression in the FDM halo mass function implies the cumulative number density of halos asymptotes to some finite value. This can be compared with the total observed abundance of galaxies above the limiting luminosity of the \citetalias{Bouwens22} data.
Thus, we can provide a more conservative constraint by assuming that only one or fewer galaxies occupy each halo on average. This is expected theoretically at the faint end \citep{Ma18} and supported empirically by stringent bounds on satellite fractions at high redshift (which are only $\mathcal{O}\left(1\%\right)$ at $z\sim5-6$ \citep{Harikane22}, and may be even smaller at the lower luminosities and higher redshifts most relevant here). 
While our previous approach uses the full shape of the UVLF and delivers tighter bounds, the cumulative UVLF technique involves fewer assumptions regarding how galaxies populate their host halos. However, in the combined \citetalias{Bouwens21}+\citetalias{Bouwens22} data set, the observed cumulative number density of halos is dominated by faint galaxies only detectable in the \citetalias{Bouwens22} lensed survey. For completeness we will also provide bounds using only the \citetalias{Bouwens21} field measurements which are more robust to systematic errors.
Although the cumulative UVLF approach has been applied in previous work \citep{Schive_2016,Menci_2017,Ni_2019}, we emphasize here its robustness to galaxy-halo connection modeling uncertainties. 

The cumulative halo mass function (CHMF) as a function of the minimum halo mass scale $M_{\rm min}$ is:
\begin{equation}
    N(>M_{\rm min}) = \int_{M_{\rm min}}^\infty n(M)\text{d}M. 
\end{equation}
Given the requirement that fewer than one galaxy resides in each halo, on average, and with $M_{\rm min}$ denoting the
minimum mass of dark matter halos hosting star formation, the halo mass function is inconsistent with the cumulative galaxy number density implied by UVLF measurements\footnote{The total is driven almost entirely by the faintest resolved bins. Including blank-field \citetalias{Bouwens21} measurements does not improve our constraint.} unless:
\begin{equation}
    N(>M_{\rm min}) \geq \sum_{\text{ B22}}\phi(M_{\rm UV}) \Delta M_{\rm UV}.
\end{equation}
As emphasized previously, unlike in CDM, the strong suppression of low-mass halos in FDM implies $N(>M_{\rm min})$ is nearly insensitive to the choice of $M_{\rm min}$ (provided $M_{\rm min}<<M_{1/2}$).

Figure \ref{fig:cumulative} compares the cumulative UVLF measurements\footnote{Let us briefly comment on how the confidence intervals on the cumulative UVLFs are estimated. Simply summing independent error bars `in quadrature' is not well justified for the case of asymmetric errorbars: in fact, this would violate the central limit theorem \citep{Barlow03}. Instead, we combine errors using an approach similar to \citet{Barlow04} (which recovers the correct form in the limit of symmetric errors). We first determine the most probable set of underlying differential UVLF values d$\phi_i=\phi(M_{\rm UV}^i)\Delta M_{\rm UV}^i$ given $\sum_i\text{d}\phi_i=\Phi$ for candidate cumulative UVLF totals $\Phi$. This is a constrained optimization problem: we maximize the total log-likelihood, $\ln L$, given the \citetalias{Bouwens22} measurement set, under the constraint $\sum_i\text{d}\phi_i=\Phi$ (also forbidding d$\phi_i < 0$) using \texttt{scipy.optimize.minimize} \citep{2020SciPy-NMeth,Byrd99}. The $2\sigma$ limit on $\Phi$ is set such that there is no set of d$\phi_i$ whose sum is $\geq \Phi$ with $\ln L > -2$.

}
at $z=5,8,9$ with CHMF predictions given CDM or $m_{22}=2,4,6$. The FDM curves asymptote to finite values for $M_{\rm min}\lesssim10^8 M_\odot$ while CDM is more sensitive to $M_{\rm min}$. The $z=8$ bin provides the most constraining power, where the measurements still provide robust detections in the faintest bins even though fewer halos have virialized at higher redshifts. 
Taking a conservative value of $M_{\rm min}=10^7M_\odot$ -- this is much smaller than the relevant suppression mass, for any $m_{22}<20$, $M_{1/2}>10^9M_\odot$ -- in integrating the FDM mass function, the $z=8$ measurements alone exclude models with $m_{22}>4.3$ $(2\sigma)$. Summing the log-likelihood across the $z=5-9$ redshift bins yields a slightly stronger $m_{22}>5.1$ $(2\sigma)$ (Table \ref{table:m22}).
For comparison, applying this method to only the blank-field dataset of \citetalias{Bouwens21} yields $m_{22}>1.5$ $(2\sigma)$.

These more model agnostic constraints are important, in part because alternative faint-end behavior (beyond our fiducial double power-law) is still allowed  \citep{Sun16,Furlanetto22,Sun23,Sipple24}. Furthermore, early {\em bright}-end UVLF from JWST at high redshift may point to a different redshift dependence at $z>10$ \citep{Mason23,Bouwens23,Mirocha23,Sipple24}, which may indicate deficiencies in the semi-empirical models. 
Therefore, this constraining method may be invaluable for initial FDM constraints using JWST data (see Section \ref{sec:jwst}).

\subsection{Modelling Enhanced Star Formation in Small Halos}
\label{sec:LM-tpl}
If star formation proceeds differently in the smallest halos, we may still be able to model this alternate behavior as a simple additional power law in our UVLF modeling procedure.
To this end, we introduce two additional degrees of freedom in our $L_c(M)$ relation, a new power law index, $p_2$, and its relevant mass scale, $M_2$, (similar to the ``shallow-slope" model of \citealt{Sipple24}). That is, we suppose $L_c(M)\rightarrow L_c(M) \times (M/M_2)^{p_2-p}$ for $M\leq M_2$ and perform a new fit with a vector of parameters $\mathbf{\theta}_2=(p,q,r,M_{\rm UV,0},\log_{10}(M_1/\textrm{M}_\odot),1/m_{22}, p_2,\log_{10}(M_2/\textrm{M}_\odot))$ with the same priors on the original six parameters and conservatively chosen priors for $(p_2,\log_{10}(M_2/\textrm{M}_\odot))$ of ([8, 14], [-3.5, 3.5]).

Sampling these eight parameters with MCMC, we find this model produces a similar constraint to the cumulative UVLF, (here $m_{22}>5.8$ at $2\sigma$) (Table \ref{table:m22}). If the very smallest halos are actually bright enough to be included in \citetalias{Bouwens22} (necessitating $\sim2-3\times$ the predicted star formation efficiency of our fiducial model), this could somewhat compensate for the fewer halos at a given mass in FDM. While this may indicate more efficient star formation at the faint-end, including additional parameters always presents a danger of overfitting. This new behavior coincidentally onsets near the edge of observational capabilities, where the errors bars are relatively large. Further progress may require improving the statistical precision of the faint-end UVLF measurements and additional simulations exploring star formation in small mass dark matter halos.

In contrast, the existence of greater scatter in the $L(M)$ relation for low mass halos, potentially due to bursty star formation (i.e., a short star formation duty cycle), could strengthen the limit on FDM particle mass. If only a fraction of the $\sim10^9-10^{10} M_\odot$ mass halos are UV bright enough to be detected in the \citetalias{Bouwens22} dataset, then some of the observed galaxies must necessarily reside in still smaller mass halos to match the observed galaxy abundance.

\subsection{Effect of Prior}
\label{sec:LM-prior}
Ideally, our main conclusions should not be strongly dependent on the precise form of the prior probability distribution adopted \citep{Foreman-Mackey12,Hogg18}. To this end, it is common to use uninformative priors motivated by the principle of indifference, but the choice of prior is always somewhat subjective \citep{Gelman17a,Gelman17b,Eadie23}. So in this section we are motivated to check the robustness of our conclusions to two alternate priors on FDM mass.

Our choice of a flat prior in $1/m_{22}$ is motivated by the explicit inclusion of CDM without requiring an infinite prior volume. It has been considered in a number of past analyses \citep{Irsic17,Banik21,Dentler22,Gandolfi22} although this choice gives more weight to lighter mass decades and depends on the reference mass scale \citep{Dentler22}.

Another common choice in the literature is a flat prior in $\log_{10}(m_{\rm FDM})$ \citep{Rogers21,Dentler22,Winch24}, or in the log of the suppression halo mass scale \citep{Nadler21,Banik21}. This gives equal weight to different regions in log-space, but only over a finite interval, and so this cannot explicitly consider CDM where $m_{22}\rightarrow\infty$. The upper bounds on these intervals are chosen taking into account experimental sensitivity, although if the likelihood is not well-peaked in the prior region this choice can somewhat influence the confidence intervals on $m_{22}$. Here we choose an upper bound of $m_{22}=100$ ($M_{1/2}\approx10^8\text{M}_\odot$) as the current UVLF data do not probe small enough halo masses to distinguish heavier FDM particle masses from CDM. With a prior of $\log_{10}(m_{22})\in[-1,2]$ (instead of our fiducial case of $1/m_{22} \in[0,2]$) our $2\sigma$ UVLF modeling constraint shifts only slightly from $m_{22}>15$ to $m_{22}>18$ (Tables \ref{table:params},\ref{table:m22}).

A third choice is a Jeffreys-inspired prior. The Jeffreys prior is attractive as it is invariant under a change of variables, being proportional to $\sqrt{|\mathcal{I}|}$ where $\mathcal{I}$ is the Fisher information \citep{Jeffreys61,Kass96,Robert09}. The Jeffreys prior also naturally tends to zero in regions where the prior would otherwise dominate (as the information vanishes where the likelihood of an experimental result is independent of $m_{22}$), here precluding the need for a choosing a hard upper boundary. We calculate an approximate version of this prior, neglecting the weak covariance between FDM mass and the other model parameters. That is, we calculate the Fisher information from derivatives of the likelihood function with respect to FDM particle mass, fixing the values of the other (weakly covariant) model parameters. 
We further assume in this calculation that the UVLF measurements are drawn from Gaussian distributions with standard deviations matching the quoted errors from \citetalias{Bouwens21} and \citetalias{Bouwens22}, and use the harmonic mean in case of asymmetric errors.
Again, our constraints only marginally shift, from $m_{22}>15$ to $m_{22}>13$ (Tables \ref{table:params},\ref{table:m22}), from there being less prior-dominated volume near CDM. Therefore we conclude that our conclusions are largely insensitive to the choice of prior. 

\section{FDM Halo Mass Function Modelling Uncertainties}

\label{sec:hmf}

Although a great deal of progress has been made recently, there is not yet a single simulation which spans the large dynamic range in scale need to resolve FDM effects near the local de Broglie wavelength, $\lesssim100$ pc, while capturing statistical behavior on cosmological scales, $\gtrsim100$ Mpc \citep{Zhang18b,Li19,Hui21,May23,Lague24}.
The halo mass function relation of Equation \ref{eq:schive} is only one approximation for how structure formation occurs in FDM. As mentioned in Section \ref{sec:method-hmf}, it is fit to $N$-body simulations of \citet{Schive_2016} with FDM initial conditions (IC) (including a suppressed initial matter power spectrum) and CDM dynamics (i.e., without accounting for wave-like evolution).
While there is evidence from simulations that this is a reasonable approximation \citep{Zhang18,May23}, there are a number of other approaches for calculating the HMF in FDM. For instance, semi-analytic predictions of the HMF that also assume FDM IC and CDM dynamics (see e.g. \citealt{Kulkarni21}). 
Alternatively, one can attempt to model dynamical FDM effects. Even in linear theory, FDM predicts scale-dependent growth that differs from CDM growth on small scales \citep{Marsh13,Hlozek15,Suarez15,Li19,Lague21}, while spherical collapse may be further modified \citep{Sreenath19,Harko19,Kurinchi-Vendhan22}.
In fact, some semi-analytic models approximately account for some of these effects \citep{Marsh13,Marsh16,Du17}. On the other hand, the results of recent relatively large wave-dynamical simulations favor HMFs fairly similar to \citet{Schive_2016} \citep{May23}, although these simulations currently model FDM masses somewhat below $m_{22}=1$. 
The larger de Broglie wavelengths at smaller FDM masses make these cases easier to simulate, yet they lie beneath the range most relevant for our current constraints.

In order to asses some of the uncertainties in the HMF models, we consider how our results change under each of two semi-analytic alternatives to the \citet{Schive_2016} fitting formula. 
These are based on the extended Press-Schechter (EPS) (also known as the excursion set) formalism \citep{Press74,Bond91,Sheth:2001dp}. This approach models the HMF by considering the first-crossing distribution for the linear density field, smoothed on various scales, to exceed a threshold barrier.
In this paradigm the number density of halos of mass near $M$ is:
\begin{equation}
    n(M) = \frac{\bar\rho_m}{M^2} f(\sigma_M,\delta_c) \left|\frac{\textrm{d}\ln\sigma_M}{\textrm{d}\ln M}\right|.
\end{equation}
Here $\bar\rho_m$ is the mean matter density per co-moving volume, $\sigma_M$ is the root-mean-square (rms) density fluctuation in linear theory of a region containing mass $M$, $\delta_c$ is the critical linear overdensity for spherical collapse, $\delta_c=1.686$, and $f$ is the multiplicity function, related to the statistics of random walks.

First, we determine our conclusions with FDM ICs following the sharp $k$-space procedure of \citet{Kulkarni21}. Then we consider FDM ICs plus dynamical suppression, approximated as an effective scale-dependent collapse threshold, $\delta_c(M,z)$, following \citet{Marsh13,Marsh16}.

\begin{figure}
  \centering
  \includegraphics[width=0.45\textwidth]{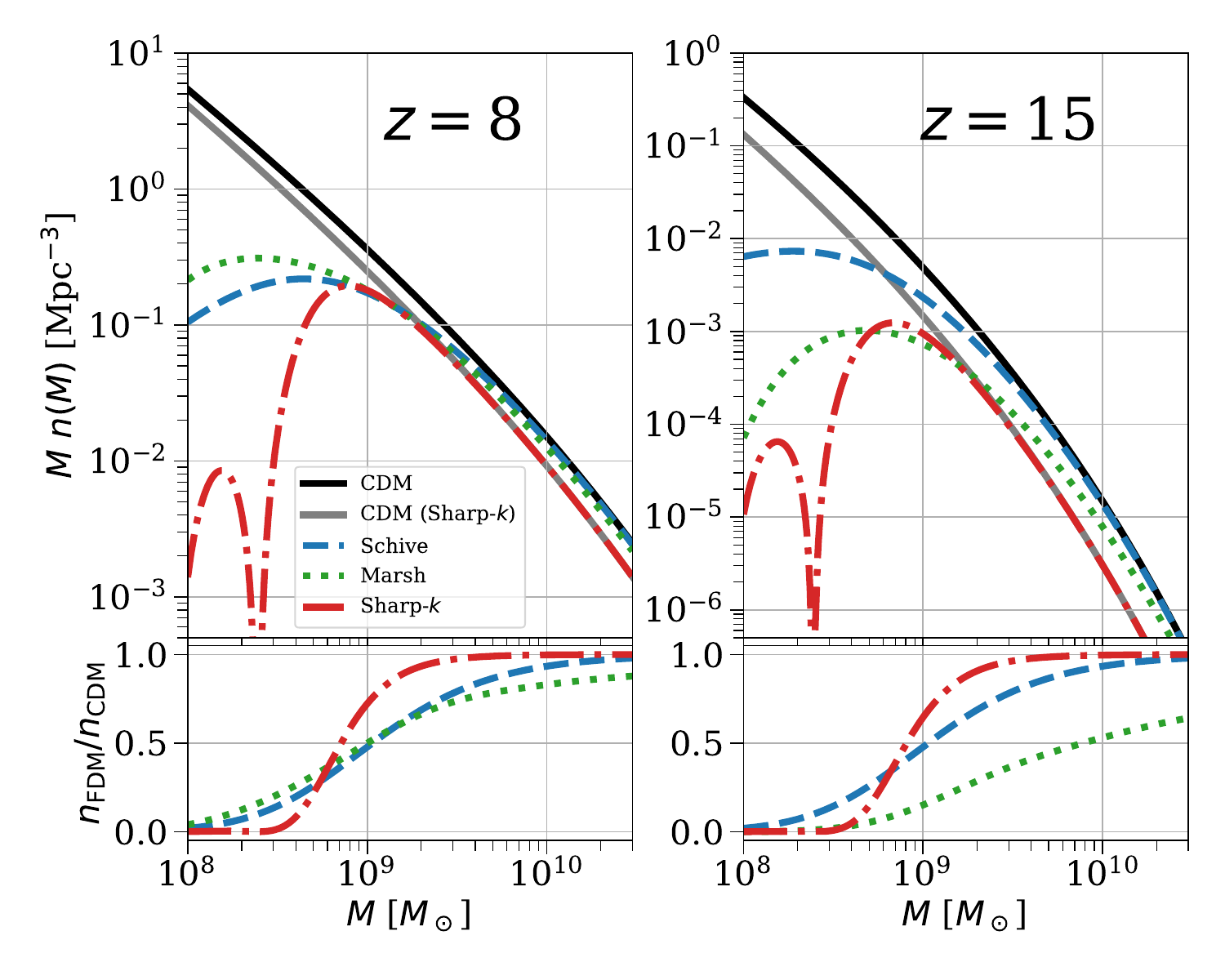}
  \caption{{\em Top}: Comparison of different FDM halo mass function models for $m_{22}=15$ at $z=8, 15$. The solid curves show CDM predictions. The fiducial case is shown in black while the gray line uses a sharp $k$-space window function (see Section \ref{sec:hmf-sk}). The Schive, Sharp-$k$, and Marsh broken colored curves represent our fiducial HMF choice, fit to simulations, (\citealt{Schive_2016}, Equation \ref{eq:schive}), an alternate analytic approach (\citealt{Kulkarni21}, Section \ref{sec:hmf-sk}), and one including the impact of dynamical suppression effects from quantum pressure (\citealt{Marsh13,Marsh16}, Section \ref{sec:hmf-marsh}, see Footnote \ref{fn:marsh-github}) respectively. {\em Bottom}: The ratio of each FDM mass function to the corresponding CDM HMF.
  }
  \label{fig:compare_mass_fns}
\end{figure}
Figure \ref{fig:compare_mass_fns} compares these two alternate halo mass function approaches to the \citet{Schive_2016} halo mass function for an FDM mass of $m_{22}=15$. The black solid curve shows the typical CDM prediction, which should be compared with the FDM predictions from the \citet{Schive_2016} and \citet{Marsh16} mass functions. The gray solid curve shows the CDM HMF calculated with the sharp-$k$ window function which should be compared to the FDM HMF calculated with the sharp-$k$ filter. 
\begin{figure}
  \centering
  \includegraphics[width=0.45\textwidth]{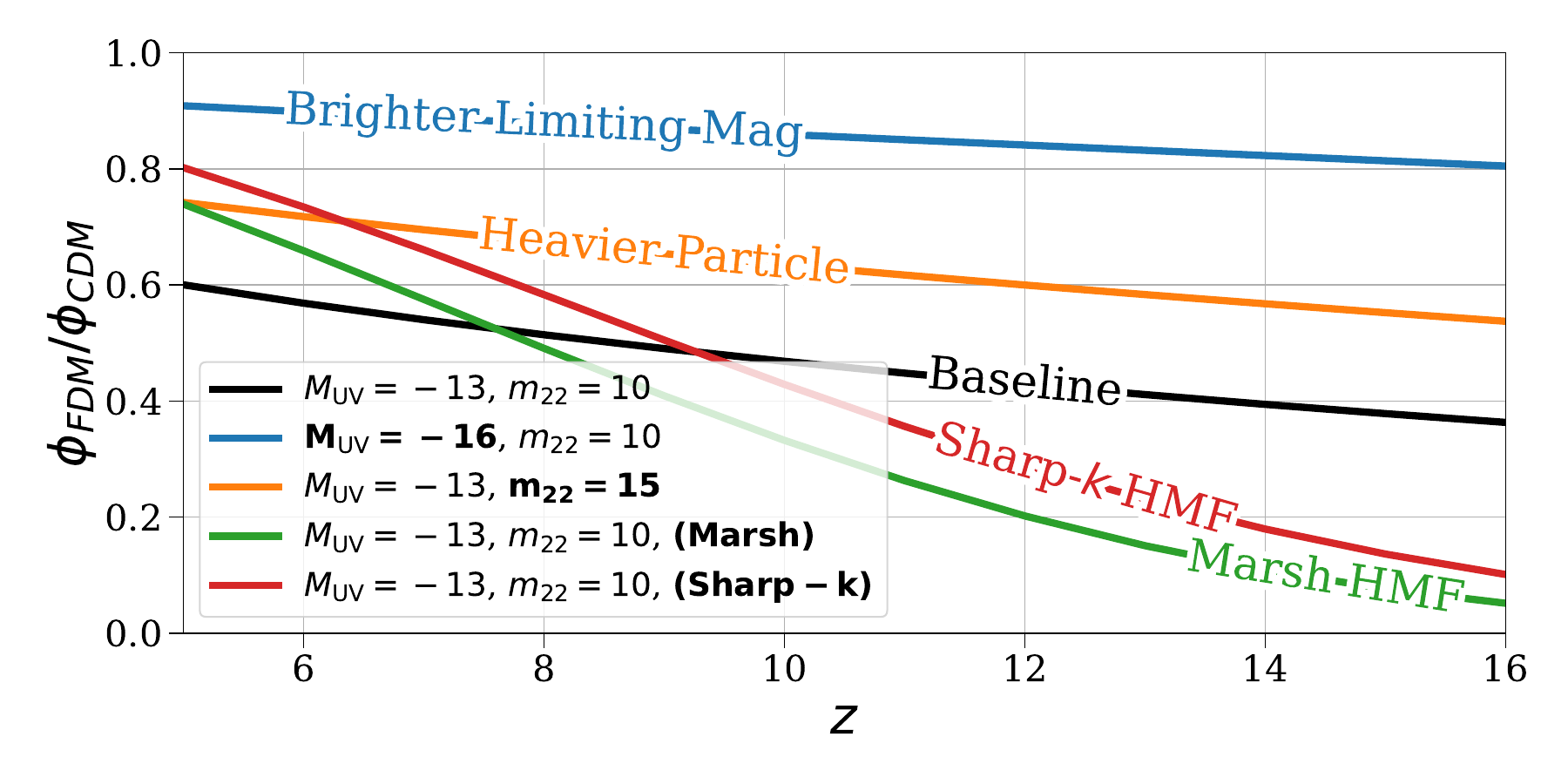}
  \caption{Ratio of FDM to CDM UVLF values in a particular luminosity bin across redshift under different scenarios. FDM is harder to distinguish from CDM if observing at brighter luminosities or if the FDM particle is heavier. The deviation at higher redshift may be slightly larger (black, orange, and blue curves) or significantly larger (red and green curves) depending on the behavior of the FDM mass function model, which is currently uncertain.
  }
  \label{fig:compare}
\end{figure}

Figure \ref{fig:compare} shows the impact of using either of these alternative HMF formulations. The $y$-axis shows the ratios of the FDM to the corresponding CDM UVLF. At a given observing magnitude and $m_{22}$ the predictions using the \citet{Schive_2016} HMF (black), sharp-$k$ space HMF (Section \ref{sec:hmf-sk}) (red) and \citep{Marsh16} HMF (Section \ref{sec:hmf-marsh}) (green) are similar for $z\lesssim10$. However, at higher redshifts (which will be probed with JWST) these alternatives predict greater suppression in the UVLF. The yellow and blue curves respectively show that FDM becomes harder to distinguish from CDM for larger $m_{22}$ or when observing only brighter galaxies.

\subsection{Sharp-$k$ Window Halo Mass Function}
\label{sec:hmf-sk}
A naive calculation of $\sigma_M$ using the FDM initial power spectrum filtered with a sharp spherical smoothing window of radius $R=\left[3M/4\pi\bar\rho_m\right]^{1/3}$ (i.e. a real-space top-hat) predicts the existence of unphysical halos below the Jeans scale \citep{Benson12,Schneider13,Urena-Lopez16,Leo18,Cedeño21,Kulkarni21}.
This results because averaging over a top-hat window in real space (or indeed a gaussian window $\Tilde{W}(kR)\propto e^{-(kR)^2}$ \citep{Schneider13}) not only includes the relevant scales but also includes a contribution from re-weighting segments of longer wavelength fluctuations \citep{Benson12,Schneider13,Kulkarni21}. This is a negligible effect in CDM but it can dominate at small scales for theories like FDM or WDM which predict a sharp truncation in power at those scales.

An alternative choice is a sharp-$k$ window function (a top-hat in $k$-space) \citep{Bond91}.\footnote{Another possibility is a \textit{smooth} $k$-space filter \citep{Leo18}.} This filter produces similar results to the real space top-hat in CDM \citep{Schneider13}, and it explicitly excises unwanted longer wavelength (smaller $k$) modes when calculating $\sigma_M$. One condition is this filter requires an appropriate assignment of real space scale, $R\sim 2.5/k$ \citep{Benson12,Kulkarni21}, and rescaling of $\delta_c\mapsto1.195\delta_c$ \citep{Kulkarni21}.
As shown in Figure \ref{fig:compare_mass_fns}, using this filter with an FDM initial matter power spectrum predicts a much sharper suppression in the HMF compared to \citet{Schive_2016}. This suppression still onsets near $M_{1/2}$ and has a weak dependence on redshift. Still, the red curve in Figure \ref{fig:compare} illustrates that this model can sometimes lead to larger differences in the UVLF (between FDM and CDM) at high redshift than in the \citet{Schive_2016}
scenario. The redshift dependence of $L_c(M,z)$ (Equation \ref{eq:L_c}, Table \ref{table:params}) implies that the same luminosity bin at higher redshift probes smaller halos. Around the FDM suppression scale, this difference is more dramatic in the sharp-$k$ case.

To investigate the difference adopting this mass function would have on our results, we performed a new MCMC sampling of our UVLF model parameters (Section \ref{sec:method-uvlf}) with the sharp-$k$ mass function. This model leads to weaker constraints with $m_{22}>8.3$ excluded at $2\sigma$. However the corresponding cumulative UVLF constraint (Section \ref{sec:LM-cumu}) is stronger, $m_{22}>7.8$ $(2\sigma)$.
These differences can both be understood as consequences of the steepness of the HMF cutoff. The \citet{Schive_2016} mass function begins to deviate from CDM predictions at higher mass scales than the sharp-$k$ model does. This is where the UVLF measurements are more strongly resolved. However, the sharp-$k$ model predicts fewer halos overall as there are almost none below the cutoff.

The difference in behavior may be due to this simple semi-analytic model missing important nonlinear effects \citep{Kulkarni21}. Alternatively, the \citet{Schive_2016} simulations may suffer from simulation artifacts (such as unremoved ``spurious halos" \citep{Wang07,Schive_2016}). In addition, the fitting formula itself may not extrapolate well outside its calibration range in particle mass ($m_{22}\sim1-3$), halo mass $M\gtrsim10^9M_\odot$, or possibly redshift \citep{Schive_2016}. Yet, current large FDM wave simulations \citep{May23} do not indicate as sharp a cutoff as predicted by the sharp-$k$ mass function.

\subsection{Scale-Dependent Collapse and Quantum Pressure}
\label{sec:hmf-marsh}

Besides the effect on the initial matter distribution, the dynamical consequences of the uncertainty principle may be important. The zero-point momenta of FDM particles implies a force which opposes gravitational confinement, known as \textit{quantum pressure}\footnote{or sometimes \textit{quantum potential}; it appears as an explicit term in the fluid formulation of the Schr\"{o}dinger Equation \citep{Madelung26,Feynman65}.} (QP), slowing growth and forbidding structure below the local Jeans scale \citep{Marsh13,Hui17,Ferreira21}. This effect is thought to be subdominant and is so far neglected in our analysis, although its relative importance is still being investigated \citep{Zhang18,May23}.

One approximate HMF modeling approach, intended to roughly account for FDM dynamical effects, is to introduce a scale-dependent collapse threshold, $\delta_c(M,z)$ into the excursion-set based models \citep{Marsh13,Bozek15,Marsh16,Du17}. If dynamical effects should hinder the formation of halos of mass $M$ at redshift $z$, a larger $\delta_c(M,z)$ leads to fewer of those kinds of halos. This can provide a sharper cutoff in the mass function, even when $\sigma_M$ is calculated using a real-space top-hat window function, and a complete suppression in halo formation below local Jeans scales.

For a specific example of this modeling approach, we consider the fitting formula from \citet{Marsh16}\footnote{\label{fn:marsh-github}There is a slight difference between the formula given in the text and the accompanying GitHub repository \texttt{WarmAndFuzzy} \citep{MarshGithub}. Including QP should lead to additional suppression. We therefore implement the GitHub version as the text version lies above \citep{Schive_2016} at $z<8$.} of a mass-dependent barrier based on the model of \citet{Marsh13}. In this case, $\delta_c(M,z)$ is 1.686 at high mass but exponentially increases near the suppression mass scale. Across $z=5-10$ this HMF matches \citet{Schive_2016} well for $M\gtrsim M_{1/2}$ but is more sharply suppressed for much smaller masses. As shown in Figure \ref{fig:compare_mass_fns}, at higher redshifts this mass function predicts an additional suppression at somewhat higher halo mass (also note the green curve in Figure \ref{fig:compare}).

For this model, we find very similar constraints as in our fiducial mass function model, with $m_{22}>14$ when modeling the UVLF and $m_{22}>5.0$ using the cumulative UVLF (both at $2\sigma$ confidence). The additional suppression in the mass function compared to \citet{Schive_2016} is only significant in regimes beyond those probed by current UVLF measurements. 

However, the differences become stark when predicting the impact on future UVLF surveys, such as from JWST. While the \citet{Schive_2016} fitting function evolves with redshift in the same way as CDM, the \citet{Marsh16} HMF predicts halos near the cutoff mass will take much longer to collapse and so will be rarer at early times. If this principle manifests in the true FDM HMF, then JWST UVLF measurements should lead to much stricter bounds on the FDM particle mass (see Section \ref{sec:jwst}). This motivates further study of FDM dynamical effects.

\section{Current State of FDM Constraints}
\label{sec:constraints}

\begin{figure*}
  \centering
  \includegraphics[width=0.75\textwidth]{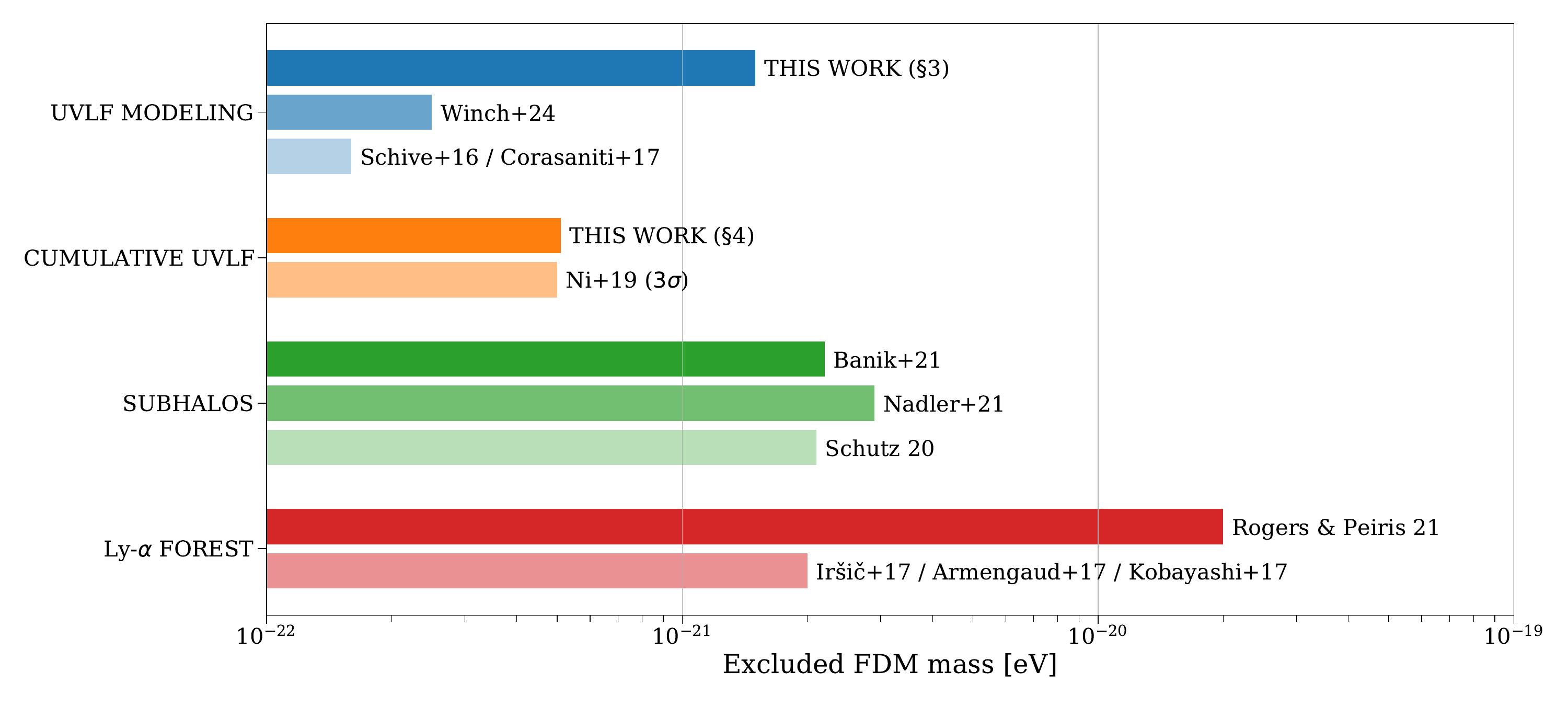}
  \caption{A selection of constraints on FDM particle mass from various methods and analyses ($2\sigma$ unless otherwise noted). (1) UVLF: this work (Section \ref{sec:results}), \citep{Winch24,Corasaniti17,Schive_2016}, (2) cumulative UVLF: this work (Section \ref{sec:LM}), \citep{Ni_2019}, (3) subhalo mass functions inferred from Milky Way satellite galaxy abundances, strong gravitational lensing, and gaps in stellar streams: \citep{Schutz20,Nadler21,Banik21}, (4) Lyman-$\alpha$ forest: \citep{Irsic17,Armengaud17,Kobayashi17,Rogers21}. See Section \ref{sec:constraints} or relevant publications for corresponding assumptions and additional details.}
  \label{fig:constraints}
\end{figure*}

Figure \ref{fig:constraints} places our competitive results in the context of a selection of other recent lower bounds on FDM particle mass in the literature. A commonality among these methods is that they all consider tracers of the amplitude of density fluctuations on sub-Mpc scales. A diverse set of approaches reaching the same conclusion reduces the influence of individual sources of error.

First, previous bounds from modeling the luminosity-halo mass relation with the UVLF \citep{Schive_2016,Corasaniti17,Winch24} are at the $m_{22}\gtrsim1-3$ level. In this work we improve these bounds by a factor of $\sim5$ by leveraging the updated \citet{Bouwens22} UVLF measurements from the gravitationally magnified galaxies in the HFF. This builds upon many years of study and refinement of the lensed UVLF analyses, including efforts from multiple groups out to $z\sim9$ \citep{Livermore17,Ishigaki18,Atek18,Bhatawdekar19,Bouwens22}.

Second, we show the similar results attained from the cumulative UVLF by \citet{Ni_2019} ($m_{22}>5$ at $3\sigma$). Their determination uses a lensed HFF UVLF analysis from \citet{Atek18} at $z=6$ compared to a cumulative HMF determined from their own FDM simulations. Their simulations predict fewer halos and thus gives a stronger constraint than the \citet{Schive_2016} HMF.\footnote{Earlier results from \citet{Menci_2017} found $m_{22}>8$ at $3\sigma$, however they relied on UVLF values from an earlier version of \citet{Livermore17}, some of which have since been revised downward \citep{Yung18,Ni_2019}.} \citet{Schive_2016} derive a bound of $m_{22} > 0.9$ from \citet{Bouwens15} cumulative UVLF data in the field, i.e. they do not consider the lensed faint-end measurements.

Third, are FDM bounds obtained from the absence of a cutoff in the inferred subhalo mass function. Recent constraints of of $m_{22}\gtrsim20-30$ have been obtained from Milky Way satellite galaxy luminosity functions \citep{Nadler21} or a combination of gaps in stellar streams and: strong gravitational lensing \citep{Schutz20} or satellite galaxy counts \citep{Banik21}.

Fourth, Lyman-$\alpha$ forest analyses, which probe the matter power spectrum by mapping the neutral hydrogen in the intergalactic medium (IGM), have determined constraints of $m_{22}\gtrsim20$ \citep{Irsic17,Armengaud17,Kobayashi17}, or more recently $m_{22}>200$ \citep{Rogers21}. These results are subject to uncertainties in the thermal state of the IGM and also to the treatment of fluctuations in the UV radiation background and in the IGM temperature-density relation, which are challenging to model in full detail (e.g. \citealt{McQuinn09}). 

While we believe that this compilation of constraints from the literature is representative, it is by no means complete. Given the broad interest in FDM, it is hard to provide a full summary of the range of techniques being considered. Additional techniques, left out of Figure \ref{fig:constraints} for simplicity, include bounds from the cosmic microwave background and large scale structure \citep{Hlozek18,Rogers23a}, black hole superradiance \citep{Stott18,Davoudiasl19}, halo core density profiles \citep{Safarzadeh20}, ultra-faint dwarf (UFD) kinematics \citep{Dalal22,Zimmermann24}, the size and age of the star cluster in the UFD Eridanus II \citep{Marsh19}, weak gravitational lensing \citep{Dentler22}, and strong gravitational lensing \citep{Laroche22,Powell23} (see \citet{Ferreira21} for additional examples and discussion). 

\section{Future JWST UVLF}
\label{sec:jwst}
JWST has already begun to make novel measurements of high redshift galaxies \citep{Harikane23,Bouwens23,Harikane24,Donnan24}. Interestingly, early results suggest a large population of UV bright galaxies even at $z\sim11-16$ --- exceeding extrapolations from the lower redshift evolution in the UVLF and the expectations of many associated models \citep{Mason23,Bouwens23,Mirocha23,Munoz24,Sipple24}. Regardless, further constraints on FDM require determining the abundance of fainter galaxies \citep{Winch24}. Ultimately, robust measurements in JWST lensing fields (see \citet{Fujimoto23} for an early example), analogous to the HFF lensing fields of HST \citep{Coe15,Lotz17,Bouwens22}, will be required to pin down the extreme faint-end. 

JWST forecasts \citep{Jaacks19,Shen23} predict detection limits in the field of $M_{\rm UV}\sim-15$. Assuming similar lensing magnification factors to those in the HFF of $\sim100$ \citep{Atek18,Bouwens22}, we expect measurements near $M_{\rm UV}\sim-10$ for JWST lensing fields. As shown in Figure \ref{fig:compare}, the ability to better probe the extreme faint-end of the UVLF may be more important than going deeper in redshift (consider the difference between the blue and black curves contrasted with the values at low and high redshifts).

For a simple forecast of JWST FDM constraints we consider the corresponding new range of halo masses that will be probed. Our HST-calibrated $L_c(M)$ model predicts that galaxies detected with $M_{\rm UV}\sim-10$ lie in halos of $M\sim4\times10^8 M_\odot$. This is a factor of $\sim5$x smaller than observed by lensed HST surveys (see Figure \ref{fig:dndM}). Since the FDM mass and HMF suppression are related by $m_{22}\propto M_{1/2}^{-3/4}$, a JWST null-detection could rule out FDM particle masses $\sim3.5$x larger (assuming similar relative errors at the faint-end as HST). It could therefore place limits of $m_{22}\gtrsim50(20)$ for the UVLF modeling (cumulative UVLF) methods. On the other hand, if JWST finds evidence for a turn-over in the UVLF at lower luminosities than probed in the HFF, this could be a signature of FDM. However, it may be difficult to determine whether such a suppression is truly caused by FDM, other alternatives to CDM, or instead reflects baryonic feedback in a CDM universe. Ultimately, one would need to combine evidence from future UVLF measurements with other techniques, such as sub-structure lensing, to be convincing \citep{Mayer22}.

At the same time, the expanded redshift range of JWST surveys will image a younger universe than HST. If FDM quantum pressure (see Section \ref{sec:hmf-marsh}) further slows structure formation above the Jeans scale, the absence of a suppression in the high-$z$ UVLF would more strongly rule out FDM (note the green curve in Figure \ref{fig:compare}). This motivates additional study of FDM dynamical effects on the HMF if one hopes to use this method to constrain $m\sim\mathcal{O}(10^{-20}$ eV). Further, faint JWST blank-field surveys could be used as a consistency check on HST lensed survey assumptions (see \citet{Bouwens22_I,Bouwens22} for potential sources of error) and decrease UVLF uncertainties in overlapping magnitude bins.

\section{Conclusion}
In this work we have explored the implications of state-of-the-art HST UVLF measurements behind foreground lensing clusters for the theory of fuzzy dark matter. To this end, we expanded the luminosity function modeling machinery used in \citet{Sipple24} to also vary over FDM particle mass. The larger de Broglie wavelengths of lighter FDM particles would inhibit the formation of small dark matter halos that host faint galaxies. Our modelling indicates that FDM scenarios with particle masses of $m<1.5\times10^{-21}$ eV predict a suppression disfavored by the UVLF data collected by HST and analyzed by \citet{Bouwens21,Bouwens22}. We further find that the cumulative UVLF disfavors $m<5\times10^{-22}$ eV even under minimal assumptions regarding how UV luminous galaxies populate their host halos.

Similarly to FDM, a warm dark matter (WDM) candidate would suppress small-scale structure (via free-streaming, \citealt{Bode01,Benson12,Schneider13,Viel13}.) A rough translation of the FDM constraints to equivalent WDM ones can be performed by matching the relevant half-mode suppression wavenumbers \citep{Marsh13,Hui17}. For example, using the thermal relic WDM half-mode wavenumber from \citet{Viel13}, the UVLF modeling (cumulative) constraints of $m_{22}\gtrsim15(5)$ translate to WDM particle masses of $m_{\rm WDM}\gtrsim3(2)$ keV. This is competitive with previous determinations using lensed HST UVLF \citep{Corasaniti17,Rudakovskyi21} ($\sim2$ keV) and early JWST UVLF measurements \citep{Maio23,Liu24} ($\sim2-3$ keV).
In future work it might be interesting to explore other scenarios, such as non-thermal WDM (e.g. sterile neutrinos \citep{Adhikari17,Abazajian17,Boyarsky19,Dasgupta21}) or interacting dark matter \citep{Tulin18,Ferreira21,Becker21,Adhikari22,Dutta24,Dave24}, which also suppress small-scale structure. Simulations exploring the differences between these theories (and the effects of baryons) will become particularly important if such a suppression is indeed found \citep{Stucker22,Shen23}. 

We also looked at a number of alternate model assumptions, which although providing similar bottom-line results highlight the need for further research. Most notably, the lack of a fully self-consistent wave simulation of FDM with cosmological box size leaves a number of open questions that will be relevant for future analyses. In the intermediate regime where FDM begins to deviate from CDM, different assumptions lead to differing predictions for halo abundances (e.g. \citet{Schive_2016,Kulkarni21,Marsh13}). 
In lieu of full simulations, research into different semi-analytic methodologies, combined with further study of and improvement in smaller-scale numerical simulations, would be useful.

Another uncertain issue relates to whether star formation proceeds similarly in CDM and FDM for intermediate mass halos \citep{Corasaniti17,Ni_2019,Shen23}.
Even under the assumptions of the CDM paradigm, the precise role of feedback in the smallest galaxies is an open question \citep{Sun16,Furlanetto22,Sun23,Sipple24}. Moreover, the surprising number of bright galaxies so far detected by JWST \citep{Harikane23,Bouwens23,Harikane24,Donnan24} may indicate that better UVLF modeling at $z\gtrsim11$ is needed \citep{Mason23,Bouwens23,Mirocha23,Munoz24}.

Nevertheless, we predict that future JWST UVLF measurements from lensing fields may increase bounds on the FDM mass by a factor of several (see Section \ref{sec:jwst}). The exact improvement here depends on a number of factors: the precise dynamics of FDM, the star formation efficiencies of small halos, and the volume and magnification factors of future lensing fields.

The cumulative number of UV photons, including the impact of galaxies too faint for JWST to observe, could also be probed by future measurements of the 21-cm signal during reionization and cosmic dawn.\footnote{In fact, the EDGES experiment's detection of the global 21-cm signal \citet{Bowman18} implies $m_{22}\gtrsim50-80$ \citep{Lidz18,Schneider18} although see \citet{Singh22} for a recent non-detection.} For instance, we find that an FDM scenario with $m_{22}\sim15$ would delay the onset of Wouthuysen–Field coupling relative to CDM by $\Delta z \sim 0.5-1$, with the precise amount depending on the FDM HMF model. The ongoing HERA experiment may eventually place bounds of up to $m_{22}\sim100$ using the 21-cm signal \citep{Munoz20,Jones21}.

It is important to develop a number of independent approaches to constrain the FDM particle mass. Each features its own sources of potential systematic error, and so many approaches converging on the same answer increases confidence in the conclusions. In Section \ref{sec:constraints} we discussed a few of these alternate methods such as the Lyman$-\alpha$ forest \citep{Irsic17,Armengaud17,Kobayashi17} or observations related to satellite galaxies \citep{Schutz20,Nadler21,Banik21}, which also constrain $m\gtrsim10^{-21}$ eV.

However, a much smaller FDM particle mass is possible if it is a subdominant fraction of the total dark matter (for instance $\lesssim15\%$ being a particle of $m\sim10^{-24}$ \citep{Winch24}). A mixed dark matter (MDM) theory with (at least) one FDM and one CDM component has theoretical motivation in string theory \citep{Arvanitaki10}. Its existence may also help resolve the $S_8$ and other tensions between probes of the matter power spectrum \citep{Amon22,Rogers23a,Rogers23b}. For these reasons, understanding the consequences of MDM on the mass function and how it might be tested is an ongoing effort \citep{Marsh13,Vogt23,Chen23,Lague24,Winch24}. 

While the nature of dark matter has remained elusive, there has been much work, in both theory and observation, in an effort to narrow-down its possible properties. For FDM in particular there continues to be exciting progress in simulations and semi-analytic models, while forthcoming JWST UVLF measurements may further constrain FDM or even provide hints of its existence.

\label{sec:conclusion}

\section*{Acknowledgements}
We thank Raghunath Ghara and Keir Rogers for helpful discussions and David J. E. Marsh for useful email correspondence. We thank the anonymous referee for a helpful report. JS and AL acknowledge support from NASA ATP grant 80NSSC20K0497. JS, AL, and DG acknowledge support from the Charles Kafuman foundation through grant KA2022-129518. GS was supported by a CIERA Postdoctoral Fellowship.

Software:
\texttt{Astropy} \citep{astropy2013, astropy2018, astropy2022}, \texttt{NumPy} \citep{numpy}, \texttt{SciPy} \citep{2020SciPy-NMeth}, \texttt{emcee} \citep{Foreman-Mackey12}, \texttt{CAMB} \citep{CAMB}, \texttt{Matplotlib} \citep{matplotlib}, \texttt{corner} \citep{Foreman-Mackey2016}, \texttt{Matplotlib label lines} \citep{labellines}.

\section*{Data availability}
The data underlying this article is available upon request. 

\bibliography{mybib}
\bibliographystyle{mnras}

\bsp
\label{lastpage}
\end{document}